\documentclass{mem}
\usepackage{natbib}\usepackage{txfonts}\usepackage{balance}
\usepackage{graphicx}
\idline{0}{0}
\begin{document}
\def\teff{$T\rm_{eff }$}
\def\kms{$\mathrm {km s}^{-1}$}
\def\gray{$\gamma$-ray\ }
\def\grays{$\gamma$-rays\ }
\title{
COMPTEL Reloaded: a heritage project in MeV astronomy
}

   \subtitle{}

\author{
Andrew \,Strong\inst{1} 
\and Werner \, Collmar\inst{1}
          }

\institute{
Max-Planck-Institut f\"ur extraterrestrische Physik,
Garching,
Germany
\email{aws@mpe.mpg.de,wec@mpe.mpg.de}
}

\authorrunning{Strong, Collmar }

\titlerunning{COMPTEL}

\abstract{
COMPTEL was the Compton telescope on NASA's Compton Gamma Ray Observatory CGRO launched in April 1991 and which was re-entered in June 2000. COMPTEL covered the energy range 0.75 to 30 MeV, and performed a full-sky survey which is still unique in this range, with no followup mission yet approved.
 This remains a major uncharted region, and the heritage data from COMPTEL are still our main source of information.
 Data analysis has continued at MPE however, since the data were never fully analysed during the mission
 or in the period following, and improvements in analysis techniques and computer power make this possible. 
 }

\maketitle{} 

\section{Introduction}
COMPTEL was the Compton telescope on NASA's Compton Gamma Ray Observatory CGRO launched in April 1991 and which was re-entered in June 2000. COMPTEL covered the energy range 0.75 to 30 MeV, and performed a full-sky survey which is still unique in this range, with no followup mission yet approved. This remains a major uncharted region, and the heritage data from COMPTEL  are still our main source of information in the MeV range. Data analysis has continued at MPE however, since the data were never fully analysed during the mission or in the period following, and improvements in analysis techniques and computer power make this desirable and possible. 

\section{COMPTEL Mission and Instrument}
\begin{figure}
\includegraphics[width=1\textwidth,clip=t,angle=0.,scale=0.5]{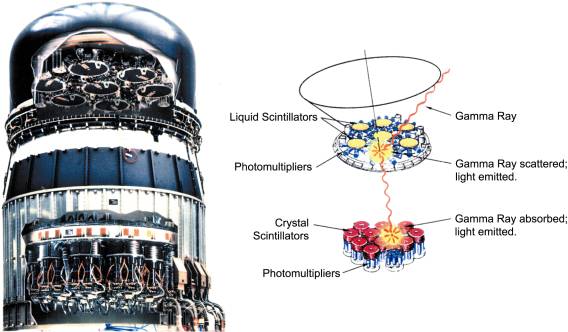}
\caption{The  COMPTEL instrument and principle of operation.
}
\label{comptel}
\end{figure}

\begin{figure}
\includegraphics[width=1\textwidth,clip=t,angle=0.,scale=0.5]{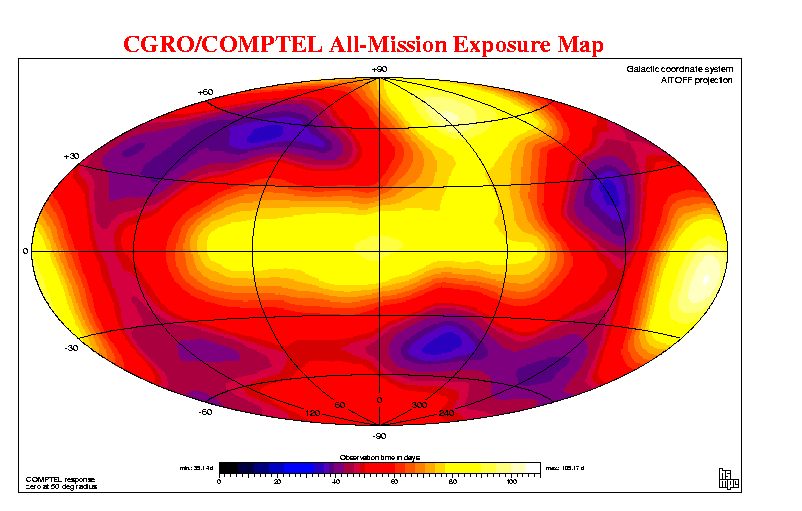}
\caption{ COMPTEL sky exposure, cm$^2$s.  Galactic coordinates, centred on l = 0, b = 0.
}
\label{exposure}
\end{figure}

COMPTEL was a double-scatter Compton telescope: incoming \grays Compton-scatter in  one of the 7 upper organic liquid-scintillator D1 detectors, and are absorbed in one of the 14 the lower NaI  D2 detectors, see Fig \ref{comptel}. 
Both D1 and D2 use photomultipliers to measure the light signal and locate the scatter position using the Anger-camera principle. The energy deposits give the Compton scatter angle according to the usual formula. Hence the incoming direction is determined to an annulus on the sky, whose width depends on the precision of the energy and position measurements. At high energies the absorption in D2 is incomplete, so the response is correspondingly broadened. The angular resolution of the Compton scatter angle is about 2$^\circ$.
The distance between D1 and D2 is 1.577m, allowing a time-of-flight (TOF) discrimination for upward-moving background \grays. 
A plastic-scintillator anticoncidence dome surrounding the instrument reduces the charge-particle background.
In addition a pulse-shape-discrimination (PSD) measurement is used for background rejection. Nevertheless, the data are background-dominated which necessitates good background-handling methods.
In its 9.7 years of operation, COMPTEL performed about 340  pointings each of roughly 2 weeks duration with field-of-view radius about 30$^\circ$, covering the entire sky, as shown in Fig \ref{exposure}.

COMPTEL also used a sophisticated multi-user analysis system called COMPASS, based on an Oracle database, which allowed full traceability of all software and  analysis operations, shared over the community of institutes involved. 
For full details of the instrument and mission see \cite{schoenfelder1993}.
Remarkably, the full mission was never analysed due to new projects taking away people-power, despite the huge investment which CGRO represented. Hence the full science potential of this major mission was never realized.

The main achievements of the COMPTEL mission included Galactic, extragalactic sources (\cite{schoenfelder2000}), $^{26}$Al maps, GRBs, solar flares, and the extragalactic diffuse \gray background. 

Meanwhile work has continued, for example
one of the strongest sources discovered by COMPTEL was in the Galactic plane at l=18$^\circ$. It was long suspected that it could be identified with the binary LS 5039 since this source has particular properties, but a proof was lacking.
Using the full COMPTEL mission data, it was possible finally to establish the identification of this source with LS5039 on the basis of the detection of its X-day orbital modulation (\cite{collmar2014}).  

\section{Recent new developments in data preparation}

\noindent
1. The TOF selection window was adapted to the newly generated TOF values (version TOF-VI), which take into account the electronic differences
of the individual D1 and D2 modules, leading to a generally narrower ToF distribution. Hence a narrower window can be defined which improves the background rejection efficiency.

\noindent
2. The PSD window was optimized as function of energy.

\noindent
3. New energy ranges were defined; the original ranges were just ad-hoc numerical  values 0.75--1, 1--3, 3--10 and 10--30 MeV (apart from $^{26}$Al which was chosen to cover the 1.8 MeV line). New ranges were defined which better avoid instrumental background lines and time variations in the response: 0.9--1.7, 1.7--4.3, 4.3--9 and 9--30 MeV. 
 
\noindent
The original analyses used a resolution in the computed Compton scatter angle (known as phibar) of 2$^\circ$. This was a compromise due to the limitations of computer resources, and leads to some degradation of the response. Now finer binning in phibar is not a problem, for example 1$^\circ$. Also the event binning and skymap binning can be finer than the original 1$^\circ$. Thus there will be no loss of potential angular resolution.

\noindent
4. Full mission data: this is a very significant development, the entire mission never having been fully analysed previously. Only the first few years were fully analyzed. CGRO underwent two orbital reboosts which affected the instrumental background; now we can analyse all data including that after the  second reboost where the background was maximum.

\noindent
5. Parts of the COMPASS data-analysis system was ported to Linux from the original Solaris environment. Thus reprocessing of the event data to the datasets needed for science can be carried out routinely with new parameters, in a fraction of the original time.

\section{New developments in imaging}
The ``classic'' maximum-entropy deconvolution method (\cite{skilling1989}) implemented in the MEMSYS5 software package (\cite{memsys5manual}) which is the basis for most of the published images (\cite{strong1999}), was adapted to use modern fast convolution-on-the-sphere methods (via spherical harmonics),  and current parallel architechtures.
 The skymaps use the  HealPix all-sky equal-area pixelisation, which can be visualized with the CDS Strasbourg Aladin interactive sky atlas,.
  Images can be generated in a fraction of the original time and with finer angular resolution. 
The data are maintained in the instrumental system to facilitate accurate response and background templates. The background template is based on a number of high-latitude observations, fitted with time-dependent scaling factor. Occasional variations in the response PSF, due e.g. to solar mode, are included on an observation basis.

\begin{figure*}
\includegraphics[width=1\textwidth,clip=t,angle=0.,scale=0.5]{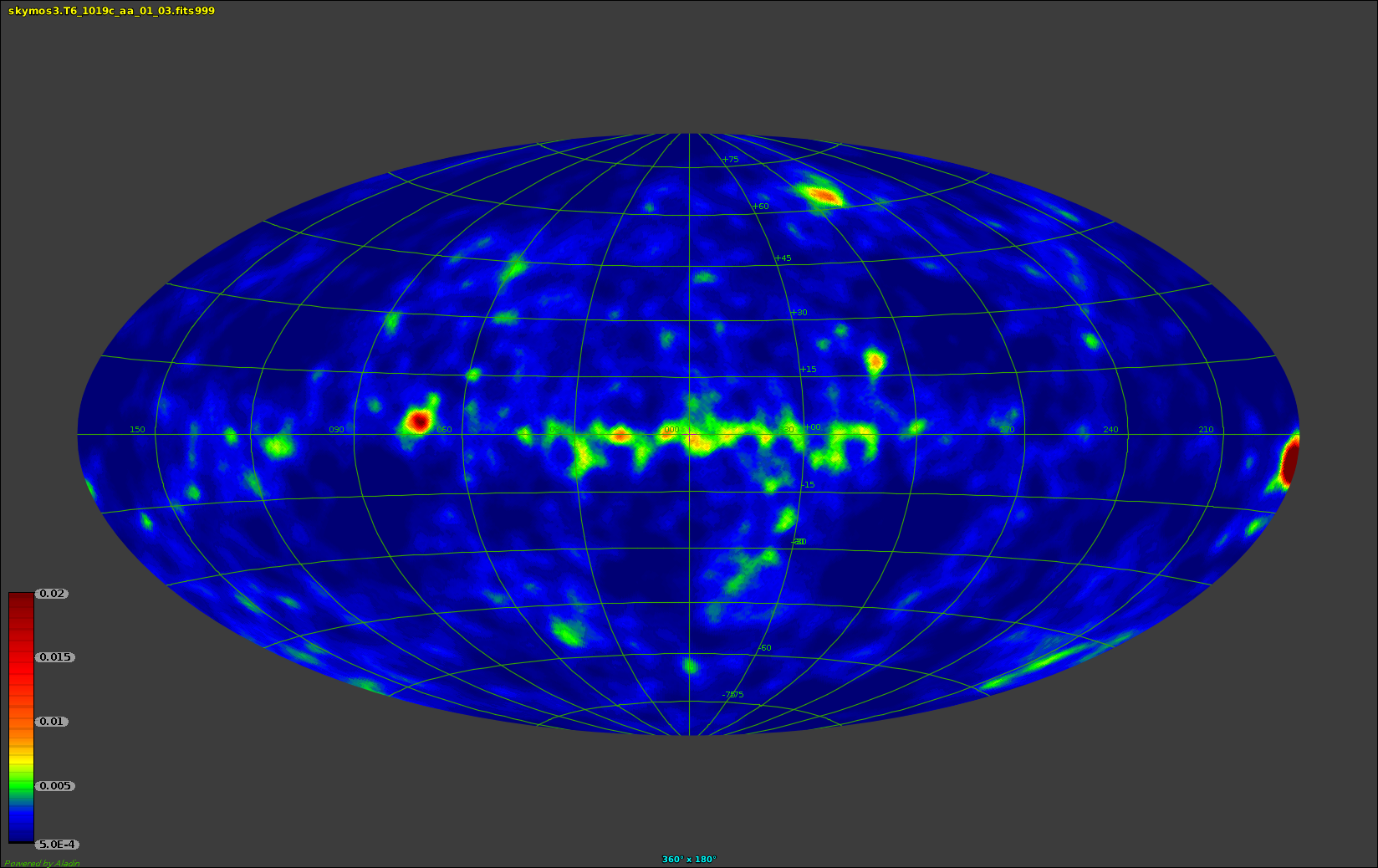}
\includegraphics[width=1\textwidth,clip=t,angle=0.,scale=0.5]{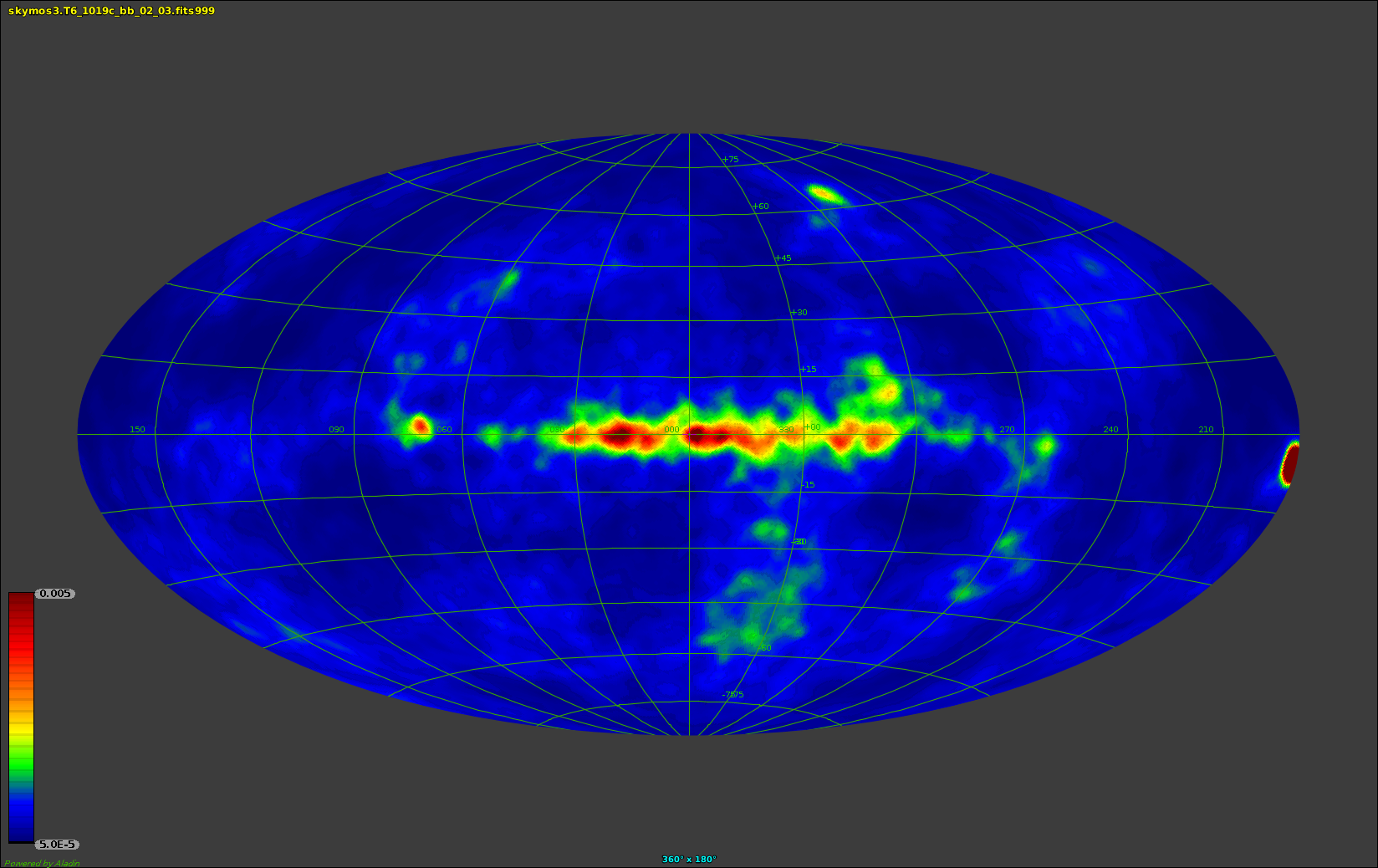}
\includegraphics[width=1\textwidth,clip=t,angle=0.,scale=0.5]{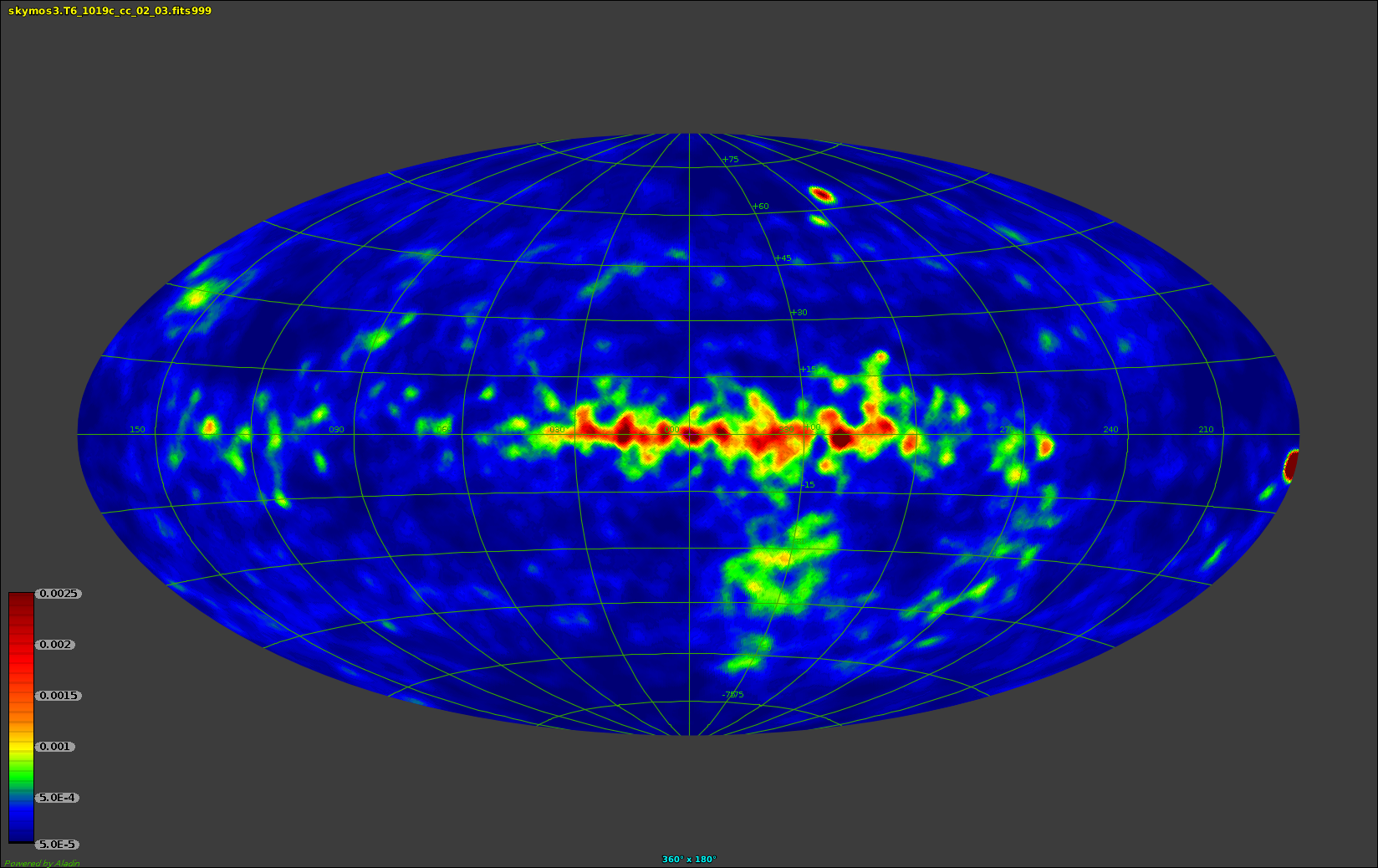}
\includegraphics[width=1\textwidth,clip=t,angle=0.,scale=0.5]{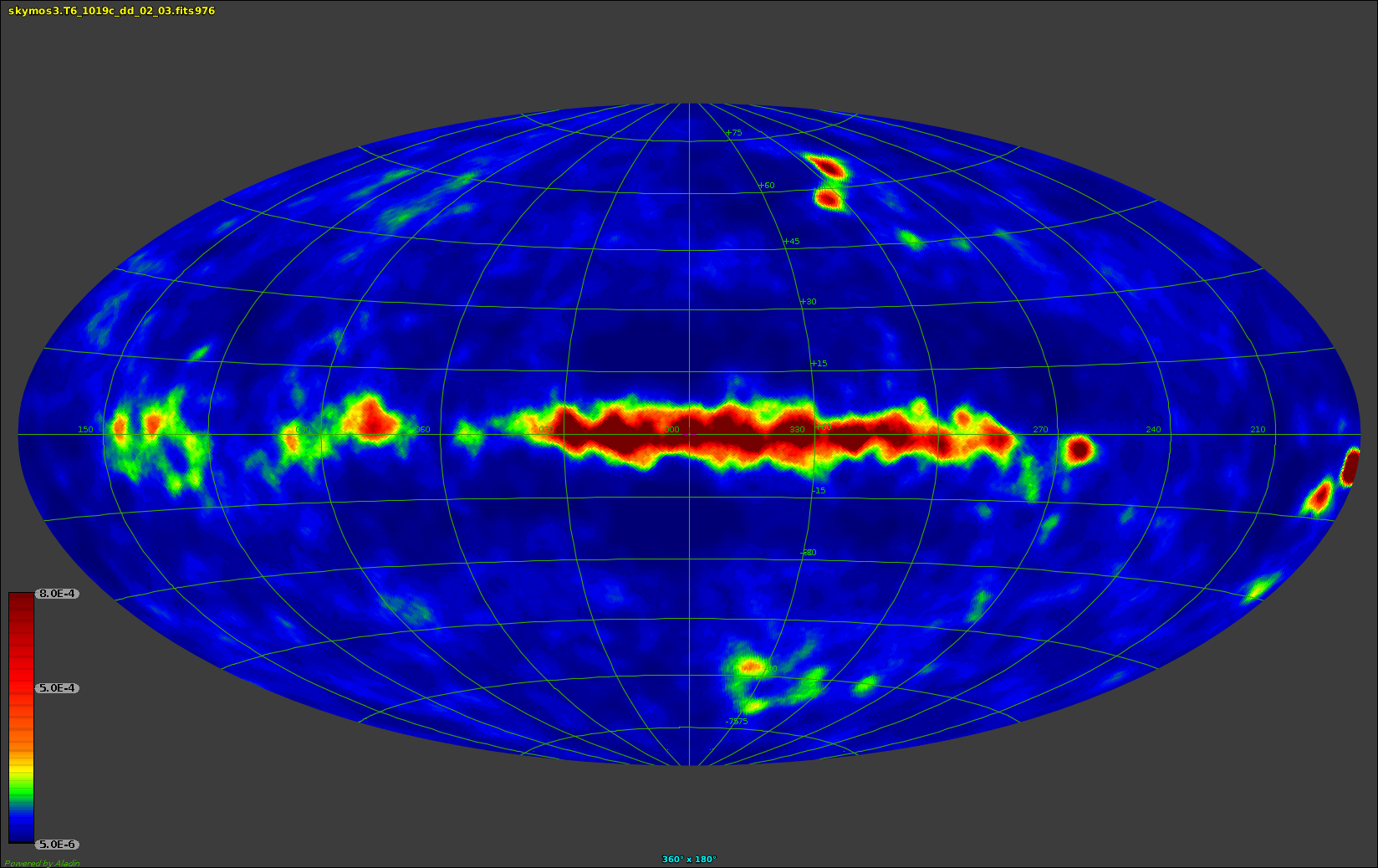}
\caption{ Preliminary COMPTEL all-sky images using the new  maximum-entropy method. Galactic coordinates, centred on l = 0,  b = 0.
Left to right, top to bottom: 0.9--1.7, 1.7--4.3, 4.3--9 and 9--30 MeV.
}
\label{skymaps}
\end{figure*}

Fig \ref{skymaps} shows preliminary maximum-entropy all-sky images in the four new energy ranges.
The Galactic plane is the most striking feature, while known sources both Galactic and extragalactic are visible. The extended feature below the plane in the fourth quadrant is thought to be a residual background effect. Detailed evaluation of these images is ongoing.

\section{New source catalogue}
The original COMPTEL source catalogue (\cite{schoenfelder2000}) contained 32 steady sources, both Galactic and extragalactic. The new full-mission data and analysis are enabling the production of a new catalogue with significantly more sources and spectral and temporal measurements.

\section{Outlook}
The skymapping will be completed and extended to include $^{26}$Al and other lines of interest. These maps will eventually be made available to the community.
Interpretation of the diffuse continuum emission in the context of cosmic-ray models (\cite{bouchet2011,strong2011}) and combined with Fermi, INTEGRAL data is foreseen.

\begin{acknowledgements}
 We thank Martin Reinecke (MPA Garching) for his assistance in adapting the maximum entropy imaging software as described in this paper.

\end{acknowledgements}

\bibliographystyle{aa}

\end{document}